\documentclass[manuscript,screen]{acmart}
\AtBeginDocument{%
  \providecommand\BibTeX{{%
    \normalfont B\kern-0.5em{\scshape i\kern-0.25em b}\kern-0.8em\TeX}}}

\setcopyright{rightsretained}
\copyrightyear{2023}
\acmYear{2023}
\acmDOI{}

\acmConference[CCAI 2023]{CHI 2023 Workshop on Child-centred AI Design: Definition, Operation and Considerations}{April 23, 2023}{Hamburg, Germany}
\acmBooktitle{CHI 2023 Workshop on Child-centred AI Design: Definition, Operation and Considerations, April 23, 2023, Hamburg, Germany} 

\usepackage{placeins}

\begin{document}

\title{Designing a Realistic Peer-like Embodied Conversational Agent for Supporting Children’s Storytelling}


\author{Zhixin Li}
\authornotemark[1]
\affiliation{%
  \institution{University of Michigan}
  \city{Ann Arbor}
  \country{USA}}
\email{zhixinli@umich.edu}

\author{Ying Xu}
\affiliation{
  \institution{University of Michigan}
  \city{Ann Arbor}
  \state{Michigan}
  \country{USA}
  \email{yxying@umich.edu}
}

\begin{teaserfigure}
  \includegraphics[width=\textwidth]{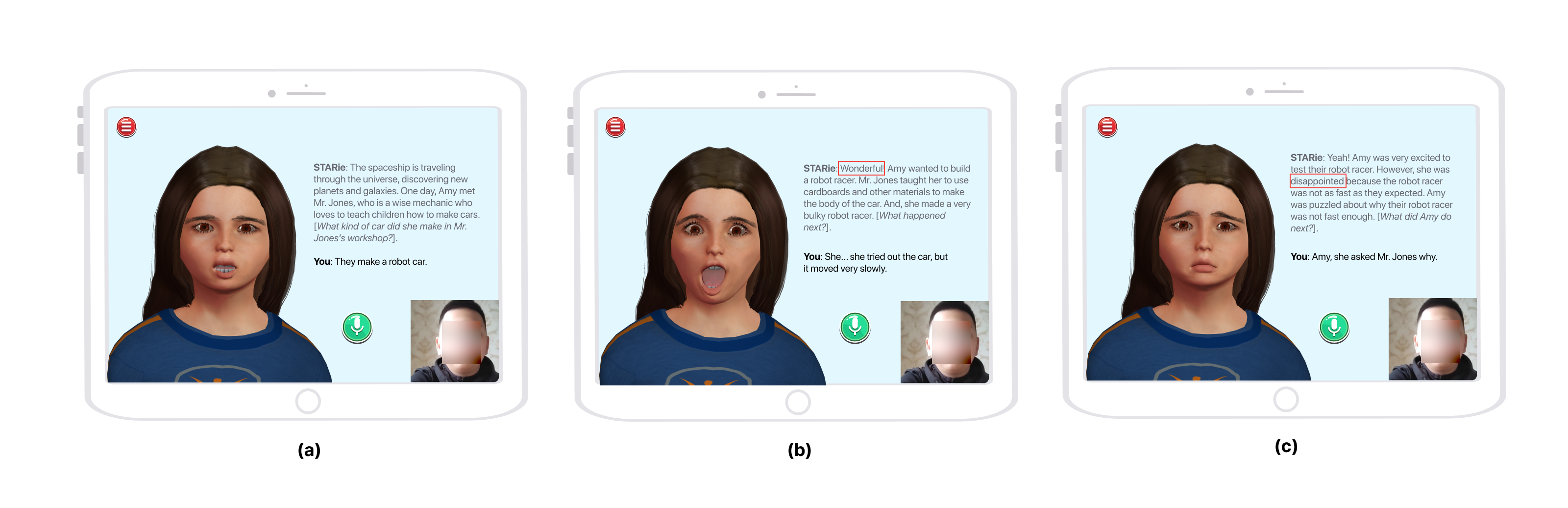}
  \caption{(a) STARie, the embodied conversational agent, tells a story with a child-like voice, appearance and real-time lip-sync and full-face animation, (b) STARie responds to the child's previous story with positive feedback and joyful facial expressions, (c) STARie empathizes with the child's response: "\emph{...but it moved slowly.}" and gives a sad facial expression and response to prompt the child to think.
}
  \label{fig:teaser}
\end{teaserfigure}


\begin{abstract}
Advances in artificial intelligence have facilitated the use of large language models (LLMs) and AI-generated synthetic media in education, which may inspire HCI researchers to develop technologies, in particular, embodied conversational agents (ECAs) to simulate the kind of scaffolding children might receive from a human partner. In this paper, we will propose a design prototype of a peer-like ECA named STARie that integrates multiple AI models – GPT-3, Speech Synthesis (Real-time Voice Cloning), VOCA (Voice Operated Character Animation), and FLAME (Faces Learned with an Articulated Model and Expressions) that aims to support narrative production in collaborative storytelling, specifically for children aged 4-8. However, designing a child-centered ECA raises concerns about age appropriateness, children’s privacy, gender choices of ECAs, and the uncanny valley effect. Thus, this paper will also discuss considerations and ethical concerns that must be taken into account when designing such an ECA. This proposal offers insights into the potential use of AI-generated synthetic media in child-centered AI design and how peer-like AI embodiment may support children’s storytelling. 
\end{abstract}





\maketitle

\section{Introduction}
Oral narrative, or storytelling skills, have been shown to be closely linked to a child's emergent literacy development, as well as their long-term language skills and well-being~\cite{CASSELL200475}. From a young age, children start engaging in storytelling and continue to develop more fluent and mature language structures as they grow older. Studies indicate that when children collaborate with a familiar conversational partner, they are better able to generate more complex stories. This is in part because they can assume shared knowledge and receive support and guidance as needed. ECAs have the potential to simulate a social partner to engage children in dialogic reading~\cite{10.1145/3563659.3563670} through asking questions, giving feedback, and adjusting questions to the developmental level of the child~\cite{10.1145/3290607.3299035}. However, most ECAs, even though they target child users, are designed to be adult partners, using adult voices and appearances. Fewer studies have explored the design of ECAs as peers, with a child-like appearance and a voice that speaks like a child. Theoretically, ECAs that take on distinct social roles, such as teachers or peers, may elicit different responses from children and may, therefore, affect their learning in varied ways. Given the great potential of utilizing ECAs to support children’s oral narrative, we should examine closely two research questions: (1) What are some of the important features for designing a peer-like ECA to support children’s storytelling? What are the potential advantages of using peer-like ECAs as compared to adult-like ECAs for children? (2) What are the ethical considerations and potential risks associated with creating child-centered ECAs, such as STARie that can mimic children's voices and appearances, and how can they be mitigated?

\section{Related Work}
The advent of generative conversational AI offers new possibilities for creatively new human-agent collaboration patterns. Generative conversational AI models, like OpenAI's GPT series, are trained on large datasets containing diverse text samples, enabling them to acquire a broad understanding of various topics, language structures, and conversational patterns. Generative conversational AI can generate context-appropriate responses based on input received from users, allowing for a wide range of applications. These include virtual assistants, conversational agents for customer service, healthcare, and even in the field of education as learning companions, language tutors\cite{10.1145/3330430.3333643}, and storytelling partners\cite{Ong2018}. 

BookBuddy is a scalable foreign language tutoring system that could automatically construct interactive lessons for children based on reading materials~\cite{10.1145/3330430.3333643}. Their pilot study results indicated that children were highly engaged while interacting with the system and preferred speaking English with their chatbot over human partners. 

Orsen is a conversational storytelling agent whose dialogue moves such as feedback, pumps, and hints which are adapted from are used to promote collaborative behavior based on identified speech acts from the child’s input~\cite{Ong2018}. The main findings have shown that most children were able to accept the virtual peer after a few trials but children tended to ignore the agent’s generic and neutral replies as they found these to be predictable.

EREN is a conversational storytelling agent that employs the OCC Emotion Model and the active listening strategy to recognize the child’s emotions and to listen to their narration of their personal stories~\cite{10.1145/3392063.3394405}. As a facilitator, EREN opens a space for discussion to help children reflect on their emotions, then transitions to a listener role to empathize while encouraging children to share events that trigger their emotions. 

StoryBuddy is a system where parents collaborate with AI in creating interactive storytelling experiences with question-answering for their children~\cite{10.1145/3491102.3517479}. A user study with 12 pairs of parents and children found that StoryBuddy was effective in providing parents with desired levels of control and involvement while maintaining children’s engagement in the storytelling process. Parents and children considered StoryBuddy useful, helpful, and likable.

Our design of the storytelling agent will be built upon the previous research and incorporate multiple design features that have been proven to be effective, including child-like voices, appearances, and full-face animation, with the goal of examining whether such design might further enhance children's experiences. 

\section{Prototype Design}
\begin{figure}[h]
    \centering
    \includegraphics[width=\linewidth]{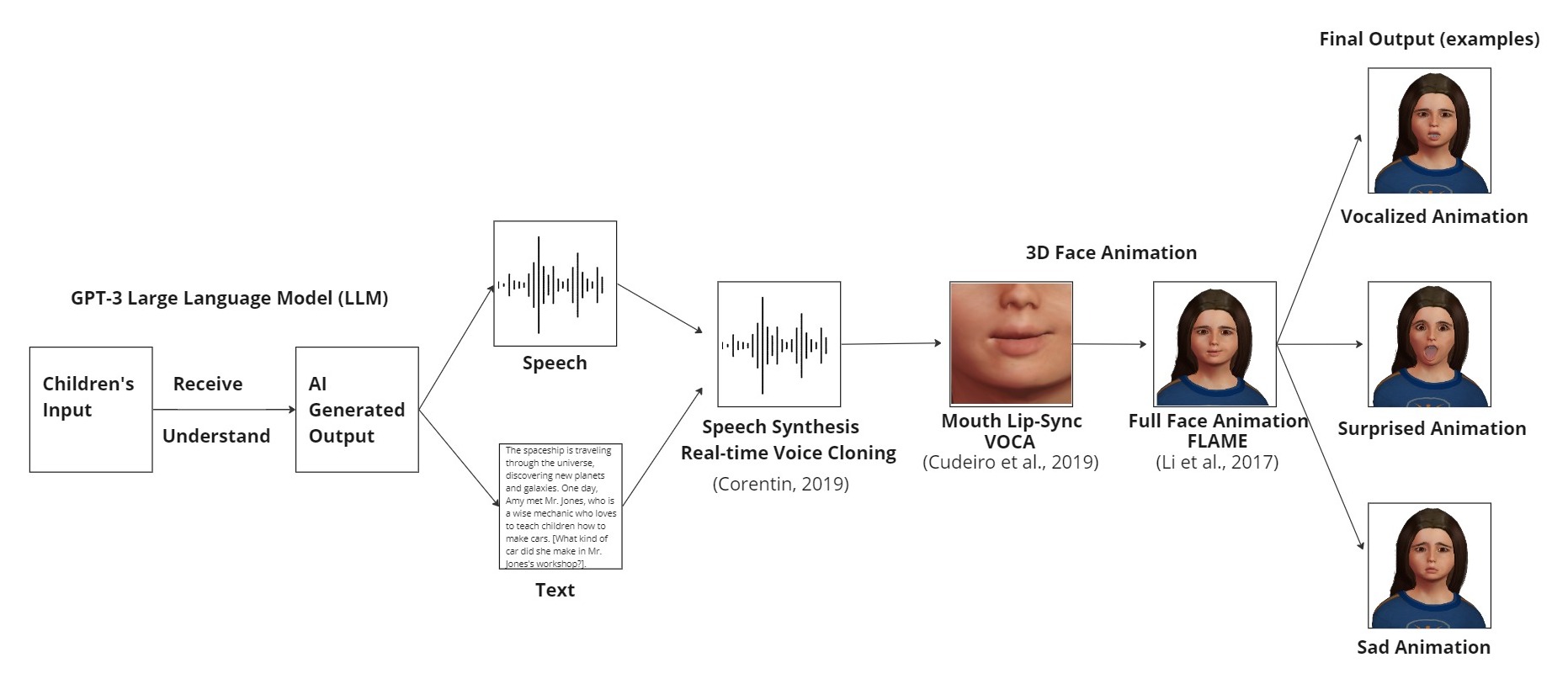}
    \caption{The pipeline that may allow STARie to generate realistic and immediate child-like voices, speech styles, facial expressions}
    \label{fig:my_label}
\end{figure}

The proposed prototype of STARie centers around the development of a 3D-modeled human girl, designed to resemble an 8-year-old. This character is created and rendered using MakeHuman and Blender, providing a relatable and age-appropriate figure for young children to co-construct stories with on a tablet-based interface (Figure~\ref{fig:teaser}). Ideally, the pipeline in action can be built on top of OpenAI's GPT-3 family of large language models that enable researchers and educators to feed STARie age-appropriate and educational effective prompts that include specific learning goals. After receiving and understanding children’ story input, STARie will immediately generate text and audio responses with a child girl voice through Speech Synthesis (Real-time Voice Cloning)~\cite{Jemine2019}, which can help us create a human child's voice-over to generate arbitrary speech through either text-to-speech or speech-to-speech in seconds. Furthermore, STARie, through VOCA (Voice Operated Character Animation)~\cite{8954000}, can take any speech signal as input and animate a variety of realistic speaking styles on her face. VOCA provides animator controls to alter the speaking style, identity-dependent facial shape, and pose (i.e., head, jaw, and eyeball rotations) during animation. While VOCA automatically synchronizes the mouth with the given voice clip, the FLAME model (Faces Learned with an Articulated Model and Expressions)~\cite{10.1145/3130800.3130813} will be used to animate the neutral face generated in the previous step. It also allows the researchers to manually customize the expressions and movements of the generated 3D face. Last, the final output can be animated with various speaking styles and facial expressions (Figure~\ref{fig:my_label}).

\section{Discussion}
\subsection{Expected mechanisms of how STARie as a peer may support collaborative storytelling}
\subsubsection{Rapport-building with children} \

Studies have shown that CAs with social interaction capabilities can enhance learning and idea
generation productivity by providing dynamic support for learners working together~\cite{5669250}. By adopting real-time voice cloning technology, STARie can use a child's voice and speech styles to interact with children. In this way, children will feel comfortable co-constructing stories with a peer-like conversational agent. Furthermore, the lip-sync and full-face animation technology, ideally, allows STARie to display lifelike facial expressions. By displaying emotions that match the story being told, STARie can help children understand the importance of emotion in storytelling and help them learn to recognize and respond to non-verbal communication, improving their social skills.

\subsubsection{Communicating educationally effective language} \

By incorporating the large language models into STARie, we plan to engineer our prompts so that STARie will be able to respond with educationally effective language. For example, STARie can be prompted to incorporate a child’s contribution to the story creation but can also be prompted to ensure that the story remains within the scope of the story outline. In addition, we also specify learning goals for STARie by providing prompts such as \emph{“By the end of this story, I (the child) will understand the concept of…”} STARie can prompt children to create characters and plots, bring their personal experiences to story development, and provide feedback and empathize with children. By interacting with STARie in this way, children are expected to develop their narrative skills through guided practice.

\subsubsection{Support children with special needs} \

It is noteworthy that the combination of these models above can be further used for children with Childhood Apraxia of Speech (CAS), a motor speech disorder that makes planning and producing the precise, highly refined, and specific series of movements of the tongue, lips, jaw, and palate that are necessary for intelligible speech difficult~\cite{morgan2018interventions}. Real-time voice cloning technology can be either used for STARie to talk with children like a peer at similar age or function as an alternative digital voice output device for children who mainly rely on text-based interaction. Furthermore, with the support of the FLAME model, STARie can display many exaggerated facial expressions together with a combination of auditory reinforcements. The movements of STARie’s lip, tongue, and tooth movements can be used to improve children with CAS learn speaking by imitation~\cite{Mencia2013}.

\subsection{The adoption of STARie in early childhood requires careful consideration and raises some ethical concerns and unsolved questions}

\subsubsection{Age appropriateness} \

Although large language models such as GPT-4 and Alpaca are capable of engaging in natural conversations with humans, they are not optimized for interacting with children. Using these models off-the-shelf to develop an agent like STARie for direct interaction with children is not ideal, as they may generate inappropriate responses or biased information. Therefore, it is crucial to investigate alternative approaches to ensure that STARie generates appropriate language. This could involve implementing mechanisms to detect inappropriate language, or using databases of child-appropriate stories to fine-tune the language models used by STARie. In this way, we can enhance children's experience and promote the development of safe conversational agents for children.


\subsubsection{Protecting privacy} \

In order to create a realistic voice clone, speech synthesis software needs to analyze and process numerous voice recordings. One of the primary concerns is the collection and storage of voice data, which could involve collecting and storing large amounts of personal and sensitive data when children are co-creating stories with STARie. Children are particularly vulnerable in this context, as they may not fully understand the implications of sharing their voice data or how it might be used in the future~\cite{doi:10.1080/1369118X.2019.1657164}.

\subsubsection{Gender choices of STARie} \


As mentioned in the previous section, the proposed interface design of STARie is a humanlike girl aged around 8 years old. However, this decision is still arbitrary. It is possible that STARie's gender representation may impact children's interaction with and perception of this agent. Past research in developmental psychology has suggested the gender of peers can influence young children's play behaviors~\cite{https://doi.org/10.1111/j.2044-835X.2011.02032.x}, and it is possible such effects would be observed in the context of child-agent interactions. Furthermore, many of the existing peer storytelling agents were designed to be gender-neutral to minimize the potential effects resulting from the gender (mis)alignment with children. Yet further research is needed to investigate whether this approach is ideal.


\subsubsection{Uncanny valley phenomenon} \

It's also crucial to consider the potential for an uncanny valley effect on children. Although research has suggested that children under the age of 9 are less sensitive to the uncanny valley effects ~\cite{Brink2019}, it is still important to create child-friendly storytelling characters as children are still developing their ability to recognize and understand facial expressions, body language, and other nonverbal expressions. Young children may struggle to distinguish between the physical and virtual aspects of the experience~\cite{10.1145/3429360.3468208}.

\section{Future Work}
STARie, presently in its conceptual stage, has the primary objective of designing and implementing a child-centered storytelling experience. Therefore, collaborations with interdisciplinary researchers in the field of child psychology, education, conversational AI (CAI), natural language processing (NLP), and computer science could enrich our research, providing a more comprehensive understanding of the applications of these existing AI models to develop our project - STARie. We also plan to incorporate a participatory design approach in our future research, which involves engaging children and stakeholders actively in the design process, ensuring that their needs and preferences are taken into account.

\section{Conclusion}
In conclusion, we proposed a design prototype of a peer-like storytelling ECA named STARie that leverages existing AI models that may ideally generate child voice and lifelike face animations, which may help improve children's narrative skills during collaborative storytelling process. However, concerns around AI-generated synthetic media, including training a large number of children's audio clips to acquire authentic child voice-overs, raise many issues that touch on the awareness of age-appropriateness, children’s privacy, gender choices of ECAs, uncanny valley phenomenon, etc. Incorporating these technologies into child-centered learning needs careful and critical consideration. They are challenges, as well as implications, that require multilateral collaboration between governments, media production businesses, software developers, researchers, educators, and parents to establish new regulatory mechanisms to protect children and promote their learning and development.

\bibliographystyle{ACM-Reference-Format}
\bibliography{CHI}

\end{document}